# Toolbox model of evolution of metabolic pathways on networks of arbitrary topology

Tin Yau Pang [a,b] and Sergei Maslov [a]

[a] Department of Condensed Matter Physics and Materials Science, Brookhaven National Laboratory, Upton, NY 11973

[b] Department of Physics and Astronomy, Stony Brook University, Stony Brook, NY 11794

## Abstract

*Background:*

In prokaryotic genomes the number of transcriptional regulators is known to quadratically scale with the total number of protein-coding genes. Toolbox model of evolution was recently proposed to explain this scaling for metabolic enzymes and their regulators. According to its rules the metabolic network of an organism evolves by horizontal transfer of pathways from other species. These pathways are part of a larger "universal" network formed by the union of all species-specific networks. It remained to be understood, however, how the topological properties of this universal network influence the scaling law of functional content of genomes in the toolbox model.

*Methodology/Principal Findings:*

In this study we answer this question by first analyzing the scaling properties of the toolbox model on arbitrary tree-like universal networks. We prove that critical branching topology, in which the average number of upstream neighbors of a node is equal to one, is both necessary and sufficient for the quadratic scaling. Conversely, the toolbox model on trees with exponentially expanding topology is characterized by the linear scaling with logarithmic corrections. We further generalize the rules of the model to incorporate reactions with multiple substrates/products as well as branched and cyclic metabolic pathways. To achieve its metabolic tasks the new model employs evolutionary optimized pathways with the smallest number of reactions. Numerical simulations of this realistic model on the universal network of all reactions in the KEGG database produced approximately quadratic scaling between the number of pathways and their regulators and the size of the network. To quantify the geometrical structure of individual pathways in this model we investigated the relationship between their number of reactions and byproduct, intermediate, and feedback metabolites.

*Conclusions/Significance:*

Our results validate and explain the ubiquitous appearance of the quadratic scaling for a broad spectrum of topologies of underlying universal metabolic networks. They also demonstrate why, in spite of "small-world" topology, real-life metabolic networks are characterized by a broad distribution of pathway lengths and sizes of metabolic regulons in regulatory networks.

## Author summary

In prokaryotic genomes the number of transcriptional regulators is known to be proportional to the square of the total number of protein-coding genes. Toolbox model of co-evolution of metabolic and regulatory networks was recently proposed to explain this scaling. In this model prokaryotes acquire new metabolic capabilities by horizontal transfer of metabolic enzymes and/or entire pathways from other organisms. One can conveniently think these new pathways coming from a "universal network" formed by the union of metabolic repertoires of all potential donor organisms. While qualitative toolbox argument does not depend on specific details of the model, the exponent characterizing this scaling can be in principle model-dependent. The question we address in this study is: how the topology of the universal network determines this exponent? We first mathematically derive the quadratic scaling for a broad range of tree-like network topologies. We then propose and study the most realistic version of the model incorporating metabolic reactions with multiple substrates/products and evolutionary optimized pathways with minimal of KEGG reactions sufficient to achieve a given metabolic task. This new model combines the quadratic scaling with interesting geometrical structure of individual pathways involving byproduct, intermediate, and feedback metabolites.

## Introduction

In prokaryotic genomes the number of transcriptional regulators is known to quadratically scale with the total number of protein-coding genes [1]. Toolbox model of co-evolution of metabolic and regulatory networks was recently proposed [2] to explain this scaling in parts of the genome responsible for metabolic functions. In this model prokaryotes acquire new metabolic capabilities by horizontal transfer of entire metabolic pathways from other organisms. One can conveniently think these new pathways coming from some "universal network" formed by the union of metabolic repertoires of all potential donor organisms. The essence of the toolbox argument [2] can be summarized as follows: as the non-regulatory part of genome of an organism (its "toolbox") grows, it typically needs to acquire fewer and fewer extra new genes ("tools") in a pathway offering it some new metabolic capability (e.g. the ability to utilize a new nutrient or synthesize a new metabolic product). As a consequence, the number of pathways and by extension the number of their transcriptional regulators grows faster than linearly with the number of non-regulatory genes in the genome. While this qualitative explanation is rather general and therefore does not depend on specific details such as topology of the universal network, the exact value of the exponent $\alpha$ connecting the number of transcription factors (equal to $N_L$ - the number of pathways or leaves of the network) to the number of metabolites in the metabolic network of an organism $N_M$, as $N_L \sim N_M^\alpha$, is in general model-dependent. In [2] we mathematically derived the quadratic scaling ($\alpha = 2$) for the toolbox model with linear pathways on a fully connected graph in which any pair of metabolites can in principle be converted to each other in just one step via a single metabolic reaction. While this situation is obviously unrealistic from biological standpoint, before present study it remained the only mathematically treatable variant of the toolbox model. The universality of the exponent $\alpha = 2$ was then corroborated [2] by numerical simulations of the toolbox model with linearized pathways on the universal network formed by the union of all metabolic reactions in the KEGG database. While the agreement between the values of the exponent $\alpha$ in these two cases hinted at underlying general principles at work, the detailed understanding of these principles remained elusive.

The question we address in this study is: how the topology of the universal network affects the scaling between $N_L$ and $N_M$? To answer this question we first consider and solve a more realistic (yet still mathematically treatable) case in which the universal metabolic network is a directed tree of arbitrary topology. While being closer to reality than previously solved [2] case of fully connected network, the toolbox model on a tree-like universal network still retains a number of simplifications such as strictly linear pathways and one substrate → one product reactions.

To make our approach even more realistic we propose and numerically study a completely new version of the toolbox model incorporating metabolic reactions with multiple substrates and products as well as branched and cyclic metabolic pathways. Furthermore, unlike random linear pathways on a universal network [2] that can be long and therefore suboptimal from evolutionary standpoint, the new model uses evolutionary optimized pathways with the smallest number of reactions from the KEGG database sufficient to achieve a given metabolic task.

### Results

#### *Toolbox model on a tree-like universal network: general mathematical description*

We will first consider the case where the universal metabolic network is a directed tree. For simplicity in this chapter we will consider the case of catabolic pathways, while identical arguments (albeit with opposite direction of all reactions) apply to anabolic pathways. The root of the tree corresponds to the central metabolic core of the organism responsible for biomass production. Peripheral catabolic pathways (branches of the tree) convert external nutrients (leaves) to this core, while the internal nodes of the tree represent intermediate metabolites. Each of metabolites is characterized by its distance $0 < d < d_{max}$ from the root of the network. The universal network has $N_M^{(U)}(d)$ metabolites at distance $d$ from the root that included $N_L^{(U)}(d)$ leaves (nutrients used in the first step of catabolic pathways) and $N_B^{(U)}(d)$ branching points corresponding to intermediate metabolites generated by more than one metabolic reaction at the next level (see Fig. 1). An organism-specific network (filled circles and thick edges in Fig. 1) at distance $d$ from the root contains $N_M(d) \leq N_M^{(U)}(d)$ metabolites composed of $N_L(d) \leq N_L^{(U)}(d)$ leaves, $N_B(d) \leq N_B^{(U)}(d)$ branching points, and $N_M(d) - N_L(d) - N_B(d)$ metabolites inside linear branches ("one reaction in-one reaction out"). For simplicity we assume that in the universal network (and thus also in any of its organism-specific subnetworks) no more than two reaction edges can combine at any node (metabolite), while the most general case of an arbitrary distribution of branching numbers can be treated in a very similar fashion.

The toolbox model specifies rules by which organism acquires new pathways in the course of its evolution. It consists of the following steps: 1) randomly pick a new nutrient metabolite (a leaf node of the universal network that currently does not belong to the metabolic network of the organism) 2) use the universal network to identify the unique linear pathway which connects the new nutrient to the root of the tree (the metabolic core) and finally 3) add the reactions and intermediate metabolites in the new pathway to the metabolic network of the organism (filled circles and thick edges in Fig. 1). One needs to only add those enzymes that are not yet present in the "genome" of the organism. Graphically it means that the new branch of the universal network is extended until it first intersects the existing metabolic network of the organism.

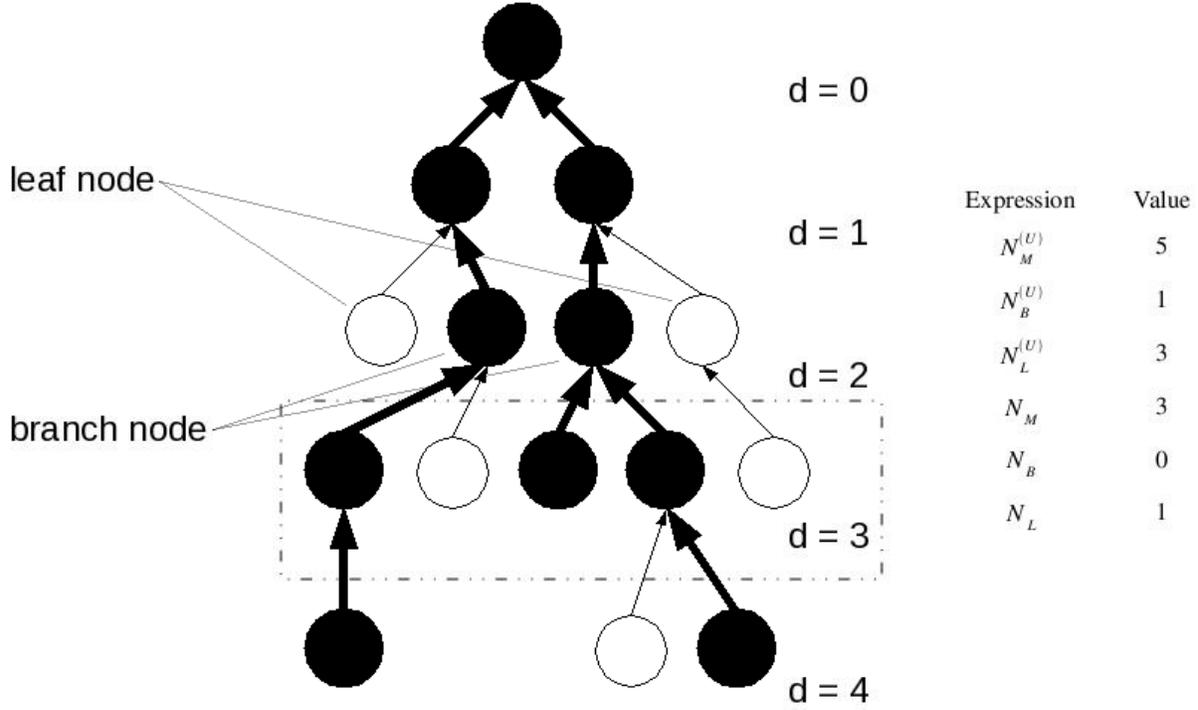

*Figure 1. An example of organism-specific metabolic network (filled circles and thick edges) which is always a subset of the universal network (the entire tree). Nodes are divided into layers based on their distance $d$ from the root of the tree. Variables $N_M^{(U)}(d)$, $N_B^{(U)}(d)$, $N_L^{(U)}(d)$ for the universal network and $N_M(d), N_B(d), N_L(d)$ for species-specific network are illustrated using the layer $d = 3$ as an example.*

Consider an organism capable of utilizing $N_L \leq N_L^{(U)}$ nutrients represented by leaves in the universal network, where $N_L = \sum_{d=1}^{d_{max}} N_L(d)$ and $N_L^{(U)} = \sum_{d=1}^{d_{max}} N_L^{(U)}(d)$. Since we assume that each nutrient utilization pathway is controlled by a dedicated transcriptional regulator sensing its presence or absence in the environment (e.g. LacR for lactose), the corresponding regulatory network would also have $N_L$ transcription factors (in the model we ignore transcription factors controlling non-metabolic functions). The non-regulatory part of the genome consists of $N_M = \sum_{d=1}^{d_{max}} N_M(d)$ enzymes catalyzing metabolic reactions (strictly speaking $N_M$ is the number of metabolites/nodes so that the number of enzymes/edges is $N_M - 1$). Quadratic scaling plots [1] shows the number of transcriptional regulators $N_R = N_L$ vs. the total number of genes in the genome (both regulatory and non-regulatory) $N_G = N_M - 1 + N_L$. However, since in all organism-specific networks $N_M \gg N_L$, the quadratic scaling between $N_R$ and $N_G$ is equivalent to $N_L \sim N_M^2$.

We further assume that due to random selection $N_L$ nutrients are uniformly distributed among all levels $d$. Therefore, the number of leaves at a given level is given by $N_L(d) = \tau N_L^{(U)}(d)$ where the fraction $\tau = N_L / N_L^{(U)}$ is the same at all levels. On the other hand the fraction $\mu(d) = N_M(d) / N_M^{(U)}(d)$ varies from level to level. It usually tends to increase as one gets closer towards the root of the tree reaching 1 for $d=0$ (the root node itself). To derive the equation for $\mu(d)$, one first notices that each of $N_M(d+1)$ metabolites at level $d+1$ is converted to a unique intermediate metabolite at level $d$. Due to merging of pathways at $N_B(d)$ branching points the number of unique intermediate metabolites at the level $d$ is actually smaller: $N_M(d+1) - N_B(d)$. To calculate $N_B(d) \leq N_B^{(U)}(d)$ one uses the fact that each of the two nodes upstream of a branching point in the universal network is present in the organism-specific network with probability $N_M(d+1) / N_M^{(U)}(d+1)$. The probability that they are both present is $\left( N_M(d+1) / N_M^{(U)}(d+1) \right)^2$ and thus the number of branching points at level $d$ of the organism-specific metabolic network is $N_B(d) = \left( \dfrac{N_M(d+1)}{N_M^{(U)}(d+1)} \right)^2 N_B^{(U)}(d)$. The intermediate metabolites together with new nutrients $N_L(d) = \tau N_L^{(U)}(d)$ entering at the level $d$ add up to the total number of metabolites at level $d$:

$$N_M(d) = N_M(d+1) - \left(\frac{N_M(d+1)}{N_M^{(U)}(d+1)}\right)^2 N_B^{(U)}(d) + \tau N_L^{(U)}(d) \tag{1}$$

This equation allows one to iteratively calculate $N_M(d)$ for all $d$ starting from $N_M(d_{max}) == \tau N_L^{(U)}(d_{max})$. We will use this equation to derive the relationship between the number of leaves and the total number of nodes first for a critical branching tree and then for a supercritical one.

### *Toolbox model on a critical tree*

The Galton–Watson branching process is the simplest process generating random trees, and we will consider its version where a node can have two, one, or zero neighbors (parents) at the previous level with probabilities $p_2$, $p_1$ and $p_0$ correspondingly. If the average number of parents $k$ equals one, then the process is referred to as critical, and if $k$ is greater than one then the process is supercritical. More generally critical and overcritical branching trees can be generated by a variety of random processes such as e.g. directed percolation [3]. While for simplicity we used the Galton-Watson branching process in our derivation below, it can be readily extended to this more general case.

The principal geometric difference between supercritical and critical trees is that in the former case the number of nodes in a layer $N_M^{(U)}(d) \sim k^d$ exponentially grows with $d$, while in a critical tree it grows at most algebraically (for the Galton-Watson critical process $N_M^{(U)}(d) \sim d^{1/2}$). The other difference is that while the critical branching process always stops on its own at a certain finite height $d_{max}$, a super-critical process will go on forever so that to generate a tree one has to manually terminate it at a predefined layer $d_{max}$. The most significant feature of a critical tree is that it has much longer branches than a supercritical one of the same size. Indeed, the diameter (the maximal height) of a random critical tree with $N_M^{(U)}$ nodes is $d_{max} \sim \sqrt{N_M^{(U)}}$ while in a supercritical tree it is much shorter: $d_{max} \sim \log N_M^{(U)} / \log k$. Thus supercritical trees (unlike their critical counterparts) have the small world property.

A random critical network where each node has at most has two parents in the previous layer is defined by $p_0 = p_2 = p \leq 0.5$. Indeed, in this case $k = 0 \cdot p_0 + 1 \cdot p_1 + 2 \cdot p_2 = 1$. In such network $N_B^{(U)}(d) = N_L^{(U)}(d) = p \cdot N_M^{(U)}(d)$ and hence the Eq. (1) can be rewritten as

$$\frac{1}{p}\left[\mu(d) - \frac{N_M^{(U)}(d+1)}{N_M^{(U)}(d)}\mu(d+1)\right] = \tau - [\mu(d+1)]^2 \qquad (2)$$

In a critical branching process $N_M^{(U)}(d) \sim d$ or $N_M^{(U)}(d+1)/N_M^{(U)}(d) = 1 + 1/d$. More generally if $N_M^{(U)}(d)$ algebraically increases with $d$, $N_M^{(U)}(d+1)/N_M^{(U)}(d)$ asymptotically approaches 1 as

$$\frac{N_M^{(U)}(d+1)}{N_M^{(U)}(d)} = 1 + \frac{const}{d} \qquad (3)$$

Thus for $1 \ll d \ll d_{max}$ $\mu(d)$ remains approximately constant and according to Eq. (2) this constant ratio $\mu$ is defined by

$$\tau = \mu^2 \qquad (4)$$

One can show (see "Calculation of the average $\mu$ in the toolbox model on a critical tree" in the Supplementary Materials) that in large critical networks the overall fraction of metabolites present in organism-specific metabolic network is very close to this stationary limit of $\mu(d)$: $N_M/N_M^{(U)} \approx \mu$.

As was explained in the previous chapter the ratio $N_G/N_G^{(U)}$ between the total number $N_G$ of metabolic-related genes in the genome of an organism and its theoretical maximal value $N_G^{(U)}$ for a genome containing the entire universal network is also given by $\mu$. Furthermore, in our model the number of leaves is equal to the number of nutrient-utilizing pathways or, alternatively, their transcriptional regulators $N_R = N_L = \tau N_L^{(U)}$. Thus like in a much simpler model of Ref. [2] the toolbox model on any critical tree-like universal network gives rise to quadratic scaling of the number of transcription factors with the total number of genes:

$$N_R/N_R^{(U)} = \left(N_G/N_G^{(U)}\right)^2 \qquad (5)$$

The geometrical properties of the universal network such as its total number of nodes/edges $N_M^{(U)} \approx N_G^{(U)}$ and number of leaves/branches $N_L^{(U)} \doteq N_R^{(U)}$ determine the prefactor of this scaling law.

### Toolbox model on a supercritical tree

For a supercritical branching process $\dfrac{N_M^{(U)}(d+1)}{N_M^{(U)}(d)} = k > 1$ and according to Eqs. (1) (see SI for the derivation) the steady state value $\mu_*$ of $\mu(d)$ satisfies

$$\tau = -\left(\frac{k-1}{p}\right)\mu_* + \left(\frac{k-1}{p}+1\right)\mu_*^2 \qquad (6)$$

Here $p = p_0$ and $k = 1 - p_0 + p_2 > 1$.

Notice that for $\tau = 0$ one has two solutions for $\mu_*$: 0 and $\mu_0 = (k-1)/(k-1+p)$. This indicates a first order phase transition in which for $\tau$ exactly at zero one has $\mu(d) = 0$, while for an arbitrary small yet positive $\tau$ the value of $\mu(d)$ asymptotically converges to $\mu_0 > 0$ for $d << d_{max}$. The number of layers over which this conversion is taking place is itself a function of $\tau$ and for small $\tau$ it is large. For exponentially growing supercritical networks and for small $\tau \ll 1$, the network average value of $\mu(d)$ defined as $\mu = N_M/N_M^{(U)}$ satisfies

$$\mu = \frac{\tau}{\mu_0} \frac{k-1}{k} \log_k \left( \frac{\mu_0}{\tau} \right) \qquad (7)$$

Note that this equation connecting $\mu$ and $\tau$ (see SI for detailed derivation) is markedly different from Eq. (6) for steady state value $\mu_*$ in middle layers.

In conclusion, while the toolbox model on a critical universal network is characterized by a quadratic scaling between $\tau$ and $\mu$ (see Eq. (4)), the same model on a supercritical, exponentially expanding universal network gives rise to a linear scaling of $\tau$ vs $\mu$ albeit with logarithmic corrections (see Eq. (7)). This difference in exponent equally applies to the scaling of the number of regulators $N_R$ vs. the total number of genes $N_G$ in the toolbox model on critical and overcritical universal network.

### *Simulation of toolbox model on the KEGG network with linearized pathways*

To test our mathematical results for a more realistic version of the universal tree we linearized pathways and reactions in the network formed by the union of all reactions in the KEGG database [4]. To this end we generated a random spanning tree on the KEGG network by the following algorithm: the metabolite pyruvate was selected as the root of the tree. We then randomly picked a metabolite located upstream of it and generated a linear pathway (tree branch) as a self-avoiding random walk on the KEGG network extended until it either merges with another pathway or reaches the root of the tree. This step was repeated until all upstream metabolites were covered. The resulting spanning tree was then used as the universal network on which we simulated the toolbox model by gradually increasing the number of pathways $N_L$ and recording the total number of metabolites $N_M$ in organism-specific metabolic networks. Our numerical simulations generated approximately quadratic scaling $\alpha = 1.8 \pm 0.1$ (see Ref. [2] and Fig. 2 of this study).

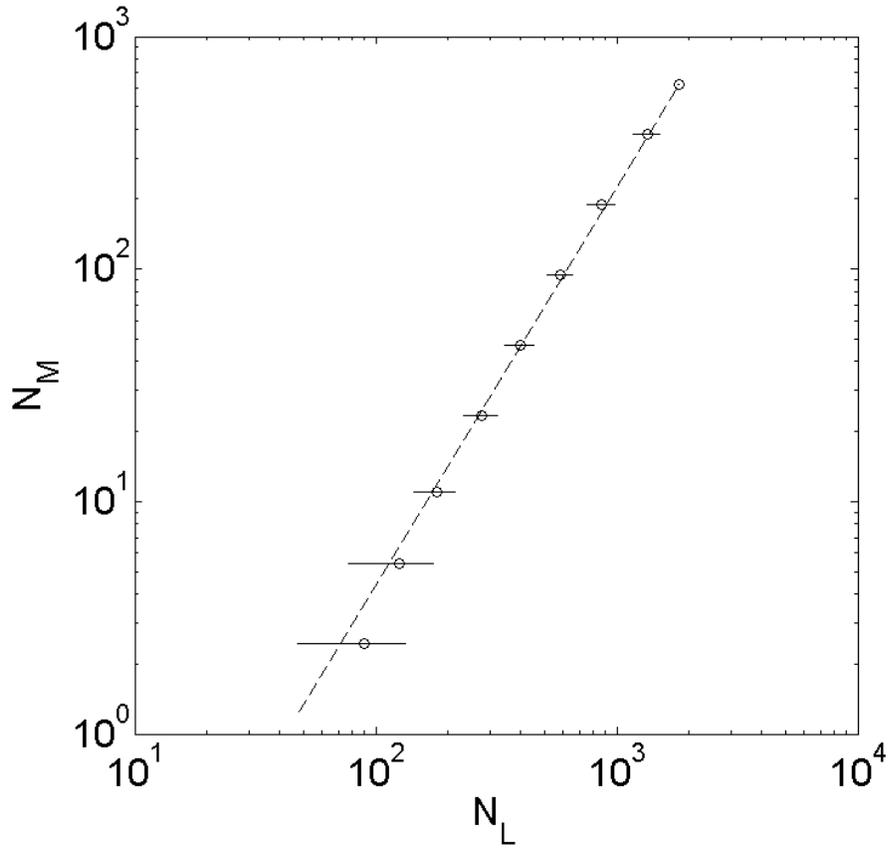

Figure 2. $N_L$ (the number of leaves equal to the number of transcriptional regulators of these pathways) vs. $N_M$ (the total number of metabolites) of the metabolic network generated by toolbox model on the KEGG universal network with linearized pathways. Solid line $N_R = N_M^{1.8}/1300$ is the best power law to the data. Error bars reflect the standard deviation of $N_M$ in 9 simulations of the model.

To better understand the origins of this scaling we investigated the topology of the underlying universal tree. The criticality of a tree is defined by the asymptotic value of the ratio $N_M^{(U)}(d+1)/N_M^{(U)}(d)$ for large $d$: for overcritical trees it reaches a limit $k > 1$, while for critical ones it converges to 1 as described in Eq. (3). Fig. 3 showing $N_M^{(U)}(d+1)/N_M^{(U)}(d)$ vs $d$ in the linearized KEGG network convincingly demonstrates its criticality. Thus the quadratic scaling between the number of transcriptional regulators and the number of metabolites in the toolbox model simulated on the linearized KEGG network is explained by the mathematical formalism described in previous chapters.

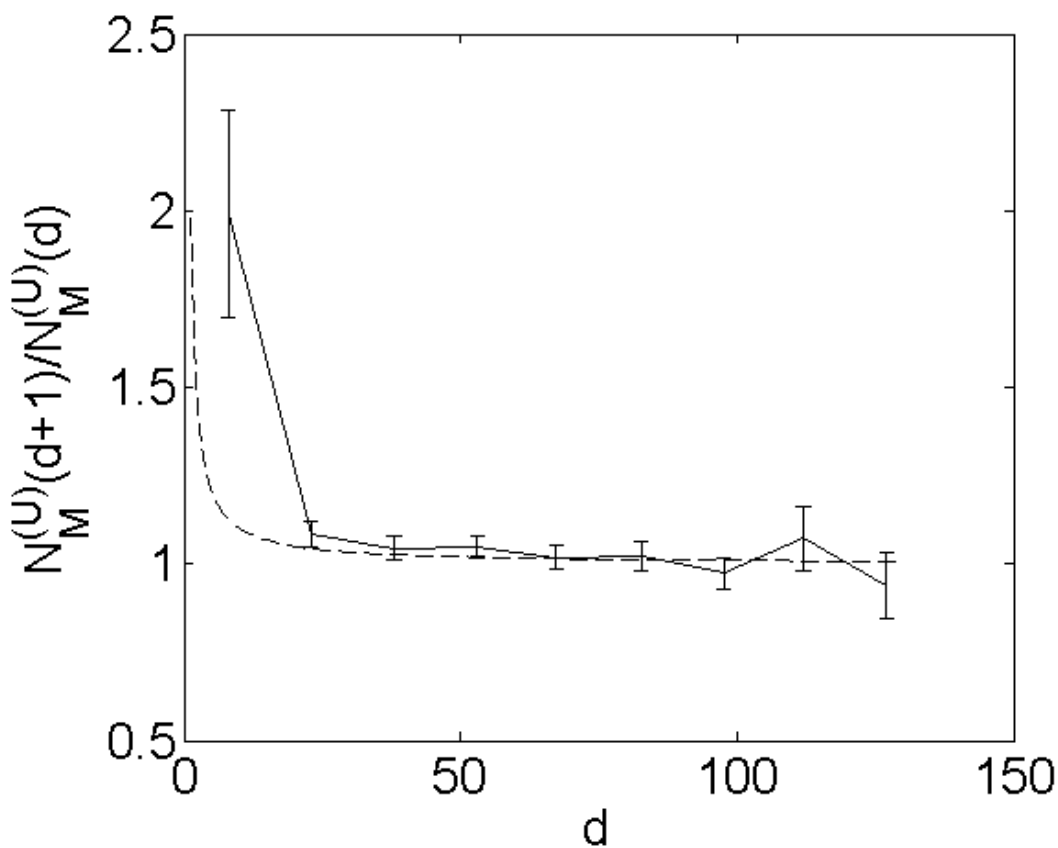

*Figure 3. The ratio of the number of metabolites at two consecutive layers,* $N_M^{(U)}(d+1)/N_M^{(U)}(d)$ *plotted as a function of the layer number* $d$ *for KEGG-based universal network with linearized pathways. Solid line: measurement, dotted line: its expected profile,* $1+1/d$*, in a critical branching tree. The error bars reflect standard deviation in different spanning trees used to linearize the KEGG network.*

In addition to using a random spanning tree to linearize the KEGG network we also tried a version using minimal paths. In this version the universal network is generated by randomly picking a metabolite and connecting it to the root of the tree (pyruvate) by the shortest path. At a first glance such "minimal path" selection appears to be reasonable from evolutionary standpoint. Indeed, evolution would favor simpler and shorter pathways in order to minimize the expenditure of resources to achieve a given metabolic goal [5] . However, the minimal paths version of linearization of the KEGG resulted in an overcritical universal network with logarithmically short branches $d \sim \log N_M^{(U)}$. As predicted for overcritical trees (Eq. (7)) the toolbox model in this case had an approximately linear scaling of the number of transcriptional regulators (leaves of branches on the network) with the total number of metabolites: the measured best fit exponent was only $1.2 \pm 0.1$ .

How does one reconcile the evolutionary pressure apparently selecting for minimal pathways with dramatically wrong scaling properties of this model? We believe that most of the ultra-short "small world" pathways generated by minimal paths on the KEGG network are unrealistic from biochemical standpoint. Indeed, highly connected co-factors often position metabolites with very different chemical formulas in close proximity to each other. For example, the KEGG reaction R00134: $\text{Formate} + \text{NADP}^+ \leftrightarrow \text{CO}_2 + \text{NADPH}$ would appear as a miraculous "one-step" conversion of carbon dioxide into formate, while the reaction R03546: $\text{CO}_2 + \text{Carbamate} \leftrightarrow \text{Cyanate} + \text{H}^+ + \text{HCO}_3^-$ would artificially link carbon dioxide and cyanate. The combination of these two reactions gives rise to equally impossible two-step path: formate → $\text{CO}_2$ → cyanate. As a consequence of such artificial shortcuts branches of the universal network linearized by minimal paths are much shorter than they are in reality. .This problem is at least partially alleviated by 1) removing unusually high-degree nodes corresponding to common co-factors such as $H_2O$, ATP, NAD in the metabolic network so that some unrealistic paths are eliminated, and also 2) using random spanning tree instead of the shortest paths. In Ref. [2] and this study (Fig. 2) we followed both of these recipes to successfully reproduce the quadratic scaling in real-life data. Still no linearization procedure could completely avoid biochemically meaningless shortcuts. In the next section we introduce and study a new considerably more realistic version of the toolbox model operating on branched and interconnected universal networks. Pathways in this version of the toolbox model satisfy the evolutionary requirements for minimal size. Yet, by properly treating metabolic reactions with multiple substrates thru les of the model prevent biochemically meaningless shortcuts and as a consequence restore proper quadratic scaling.

***Toolbox model on KEGG network with branched pathways and multi-substrate reactions.***

Real metabolic reactions routinely include multiple inputs (substrates) and multiple outputs (products) (see Table 1A-B for statistics in the KEGG database). Furthermore, metabolic networks often have two or more alternative pathways generating the same set of end-products from the same set of nutrients. Both these factors result in metabolic networks that are branched and interconnected. Here we propose and simulate a more realistic version of the toolbox model. The most prominent feature of the new model of pathways is the "AND" function acting on inputs of multi-substrate reactions. It reflects the constraint that a reaction cannot be carried out unless all its substrates are present.

The new version of toolbox model simulates addition of anabolic pathways aimed at production of new metabolites from those the model organism can currently synthesize (its current metabolic core). The new pathways are *optimal* in the sense that they contain the smallest number of reactions necessary to synthesize the desired end-product. As for previous versions of the toolbox model, one can modify the rules of this model to apply to catabolic pathways but for simplicity we will limit the following discussion to anabolic pathways. The rules of the new model are:

1. At the beginning of the simulation, the model organism starts with a "seed" metabolic network consisting of 40 metabolites classified by the KEGG as parts of central carbohydrate metabolism, plus certain "currency" metabolites such as water, ATP and NAD. It is assumed that the organism is able to generate all of these metabolites by some unspecified catabolic pathways.

2. At each step a new metabolite that cannot yet be synthesized by the organism is randomly selected from the "scope" (defined in Ref. [6]) of metabolites that are in principle reachable from the present core.

3. To search for the minimal pathway that converts core metabolites to this target we first perform the "scope expansion" [6] of the core until it first reaches the target. In the course of this expansion reactions and metabolites are added step by step (or layer by layer). Each layer consists of all

KEGG reactions that have all their substrates among the metabolites in the current metabolic core of the organism (light blue area in Fig. 4) and those generated by reactions in all the previous layers. (see Fig. 4 for an illustration).

4. Next we trace back added reactions starting from the target and progressively moving to lower levels. One starts by finding the reaction responsible for fabrication of the target metabolite and adding it to the new pathway (if several such reactions exist in the last layer we randomly choose one of them). In case of multi-layer expansion process some substrates of this reaction are not among the core metabolites (otherwise this reaction would be in the first layer). One then goes down one layer and adds the reactions fabricating these missing substrates. This is repeated all the way down to the first level of the original expansion. The resulting pathway includes the minimal (or nearly minimal) set of reactions needed to generate the target metabolite from the current metabolic core of the organism. Starting from the next step of the model the target and all intermediate metabolites become part of the metabolic core. Genes for enzymes catalyzing these new reactions are assumed to be horizontally transferred to the genome of the organism. The newly added metabolic pathway is assumed to have a dedicated transcriptional regulator so that the number of transcription factors in our model is always equal to the number of pathways or their target metabolites.

5. Steps 1 – 5 are repeated until metabolic network of the organism reaches its maximal size. At this stage it includes the entire scope [6] of the starting set of metabolites in step 1.

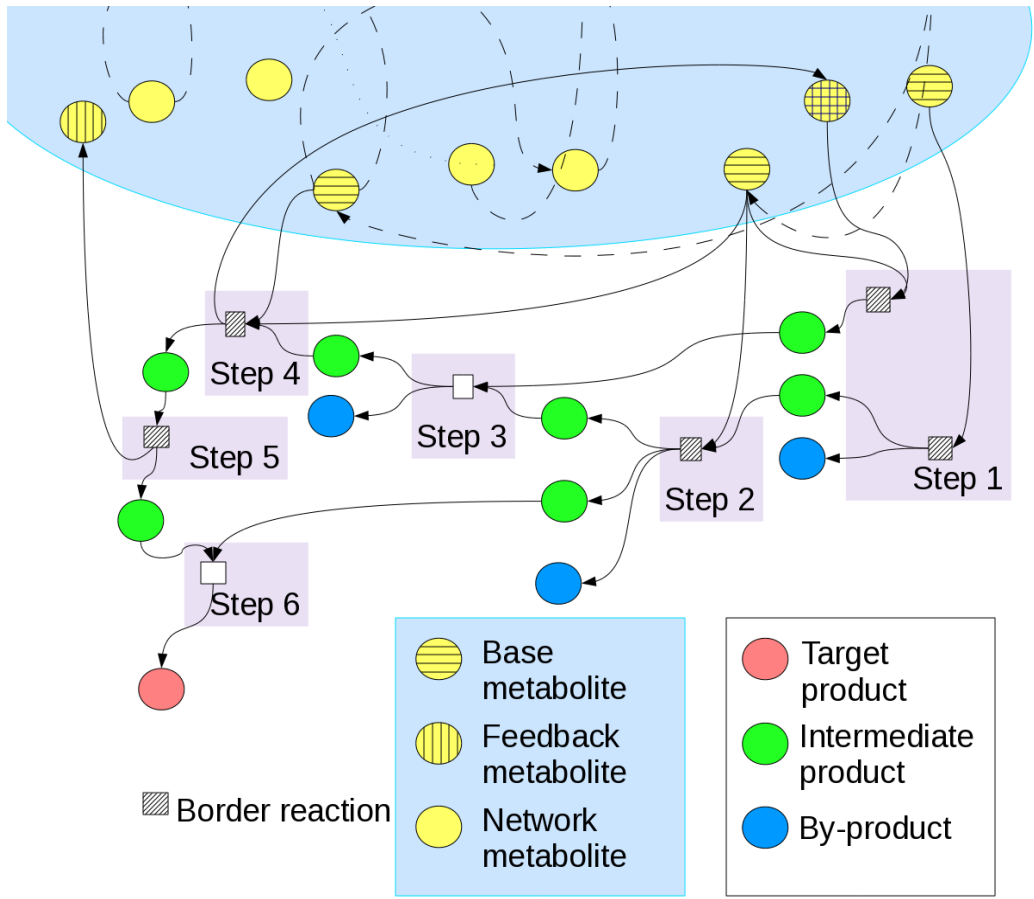

*Figure 4. Diagram of a new pathway added to the metabolic network of the organism explaining different types of metabolites and reactions. Reactions (squares) in the added pathway use base substrates (yellow circles with horizontal shading) from the metabolic core of the organism (light blue area) to produce the target metabolite (the red circle). Added pathway generates intermediate products (green circles) as well as byproducts that are not further converted to the target (blue circles). Products of some reactions feed back into the metabolic core (yellow circles with vertical shading). Reactions are labeled with expansion steps at which they were added to the pathway.*

Numerical simulation of this model shows that the number of transcriptional regulators scales with the number of metabolites with power $\alpha = 2.0 \pm 0.1$ (Fig. 5). This is consistent with quadratic

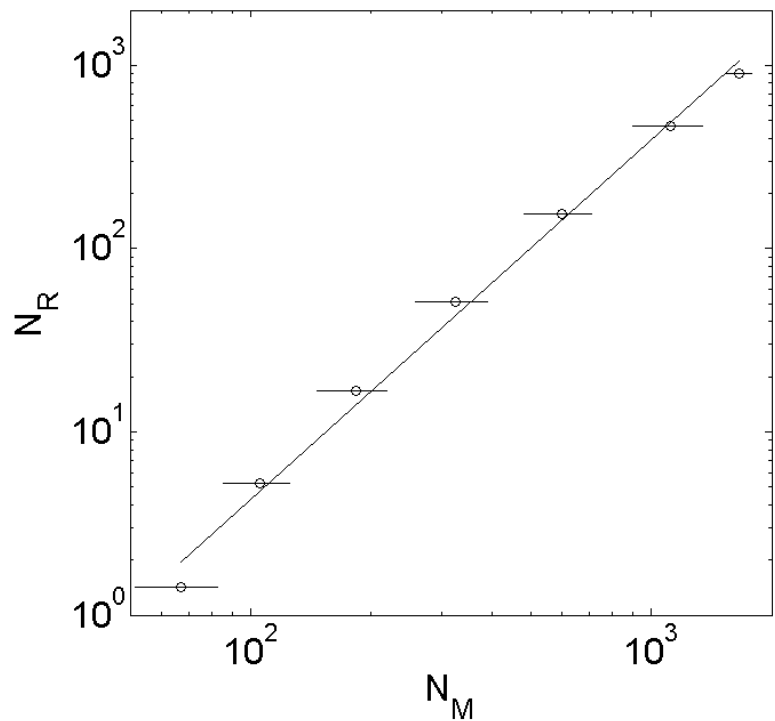

*Figure 5. The scaling between the number of pathways $N_R$ and the number of metabolites $N_M$ in metabolic networks generated by the toolbox model with branched pathways and multi-substrate reactions. Solid line with slope 2 is the best power law fit to the data. Error bars reflect the standard deviation of $N_M$ at a given value of $N_R$ in 9 realizations of the model.*

The mathematical formalism derived in the previous chapters is limited to tree-like universal networks and thus does not directly apply to the new model. Nevertheless, one generally expects the quadratic scaling to be limited only to critical, "large world" networks in which organisms with small genomes initially tend to acquire sufficiently long pathways. As noted before, from purely topological standpoint the KEGG network has a "small world" property making long pathways unlikely. It is important to check if the realistic treatment of multi-substrate reactions did in fact restore the "large world" property and criticality to the KEGG universal network by increasing the minimal number of steps required for connecting target metabolites to the metabolic core. To quantify the criticality of the expansion process as before we use the ratio $N_M^{(U)}(d+1)/N_M^{(U)}(d)$ where $N_M^{(U)}(d)$ denotes the number of metabolites reached at step $d$ of the scope expansion starting from the initial seed subset of metabolites. As in the case of critical branching trees this ratio asymptomatically converges to 1 thus confirming the criticality of the scope expansion process. The mere existence of ~40 steps in this process (the x-axis in Fig. 6) can serve as evidence in favor of "large world" character of the KEGG universal network characterized by the existence of long pathways.

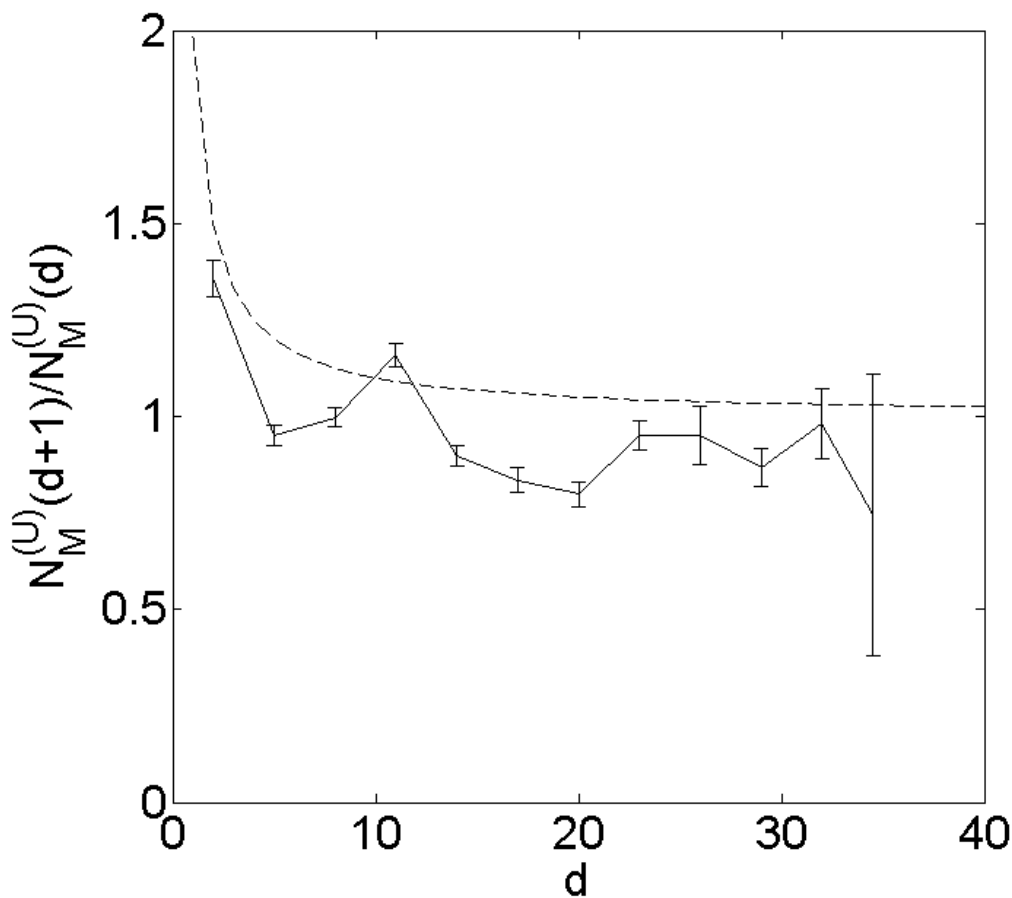

*Figure 6. The ratio $N_M^{(U)}(d+1)/N_M^{(U)}(d)$ of the number of metabolites at two consecutive layers of the scope expansion process plotted versus the layer number $d$. Scope expansion was performed*

*for the universal network consisting of all KEGG reactions. The dashed line is the mathematical expectation of the same curve in a critical branching process.*

### *Geometrical properties of branched pathways in the model.*

Unlike linearized pathways in the original version of the toolbox model [2], branched pathways in the more realistic model from previous chapter are interesting objects in their own right. We identified several geometrical properties of these pathways (see Fig. 4 for illustration) quantifying their position relative to the core network to which they were added: 1) $n_{\text{border rxn}}$ – the number of added reactions that are connected (as a substrate or a product) with at least one metabolite in the core, 2) $n_{\text{base}}$ – the number of metabolites in the core that serve as substrates to reactions in the added pathway, 3) $n_{\text{feedback}}$ – the number of core metabolites that are products of reactions in the new pathway, 4) $n_{\text{byproduct}}$ – the number of final metabolic products of the added pathway that are neither core metabolites nor the target, 5) length – the number of steps (layers of the scope expansion process) it takes to transform core metabolites into the target product. Fig. 4 illustrates the definition of these parameters while Fig. 7 and Fig. 8 plot these parameters as a function of $n_M$ (the number of metabolites in the added pathway) or $n_{rxn}$ (the number of reactions in the added pathway).

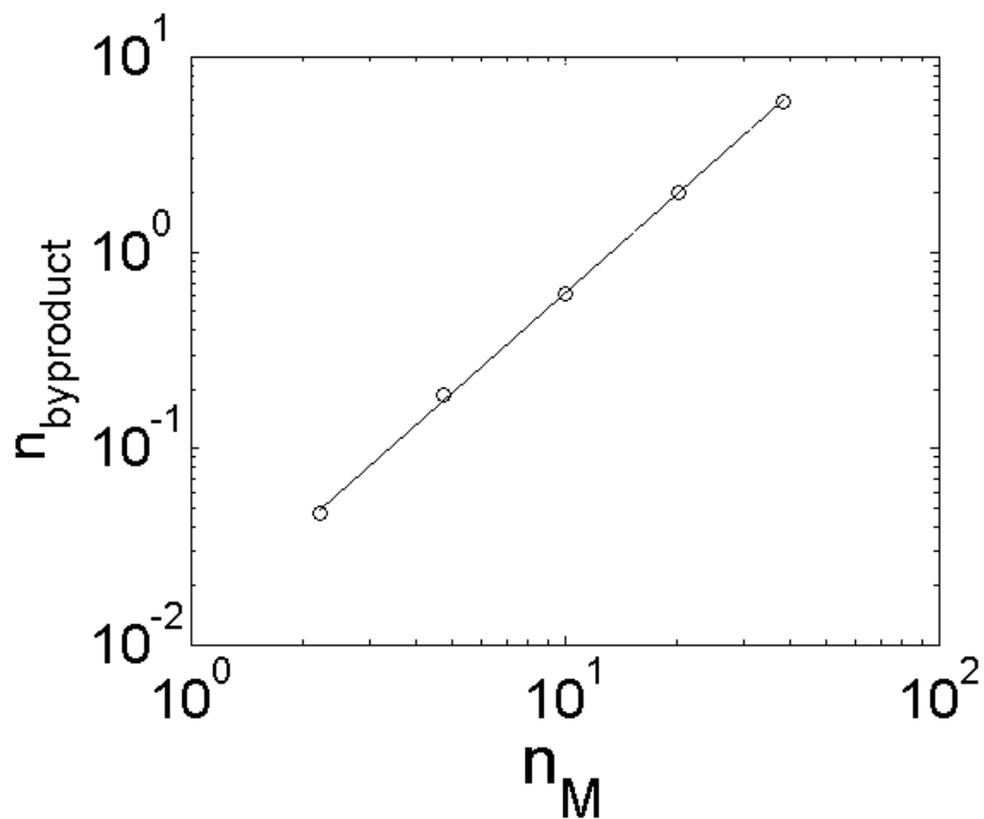

Figure 7. Faster-than-linear scaling of the number of byproducts, $n_{byproduct}$, and the total number of metabolites, $n_M$, in individual branched pathways illustrated in Fig. 4. Data for individual pathways were logarithmically binned along the x-axis. Hence y-axis can be and are below 1 due to pathways with 0 byproducts. Solid line with slope 1.7 is the best fit exponent in this plot.

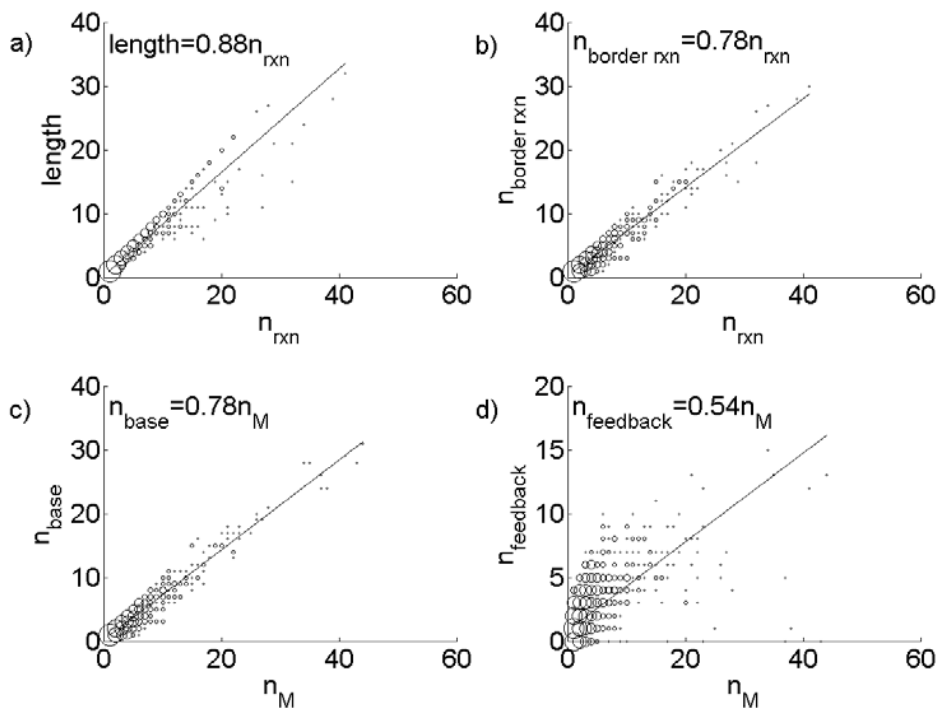

Figure 8. Approximately linear relationship between a) pathway's length and its number of reactions $n_{rxn}$, ) the number of border reactions, $n_{border\ rxn}$, and the total number of reactions, $n_{rxn}$, c) the number of base metabolites, $n_{base}$, and the total number of metabolites, $n_M$, d) the number of metabolites receiving feedback, $n_{feedback}$, and the total number of metabolites, $n_M$. These different geometrical properties of individual pathways are illustrated in Fig. 4. Sizes of circles are proportional to the logarithm of the number of discrete (x,y) pairs contributing to this point.

Approximately linear relationship between $n_{border\ rxn}$ vs. $n_{rxn}$ (Fig. 8a) suggests that added pathways tend to be located at or near the surface of the core metabolic network of the organism. Most of reactions in these pathways use metabolites from this core network either as substrates ($n_{base}$) or as products ($n_{feedback}$). On the other hand, the fact that the number of steps in a pathway (its length) constitutes a good fraction of its overall number of reactions $n_{rxn}$ implies that, in spite of these numerous "shortcut" connections to the core, added pathways remain very thin and essentially linear. That is to say, these pathways tend to work as a single "conveyor belt" sequentially converting intermediate products into each other instead of having two or more parallel "processing lines" and assembling final products of these lines only at final stages of the pathway. This finding provides an intuitive reason why models with branched and linearized pathways have similar scaling properties. One can argue that this is because pathways in both models are essentially linear. Yet, in spite of their linearity and optimality (each has the smallest number of reactions to generate the target from the core) added pathways in the new version of the model are very different from shortest paths on the universal network. As illustrated in Fig. 9 the average pathway length is several times longer than the geometrically shortest path between the target and the core.

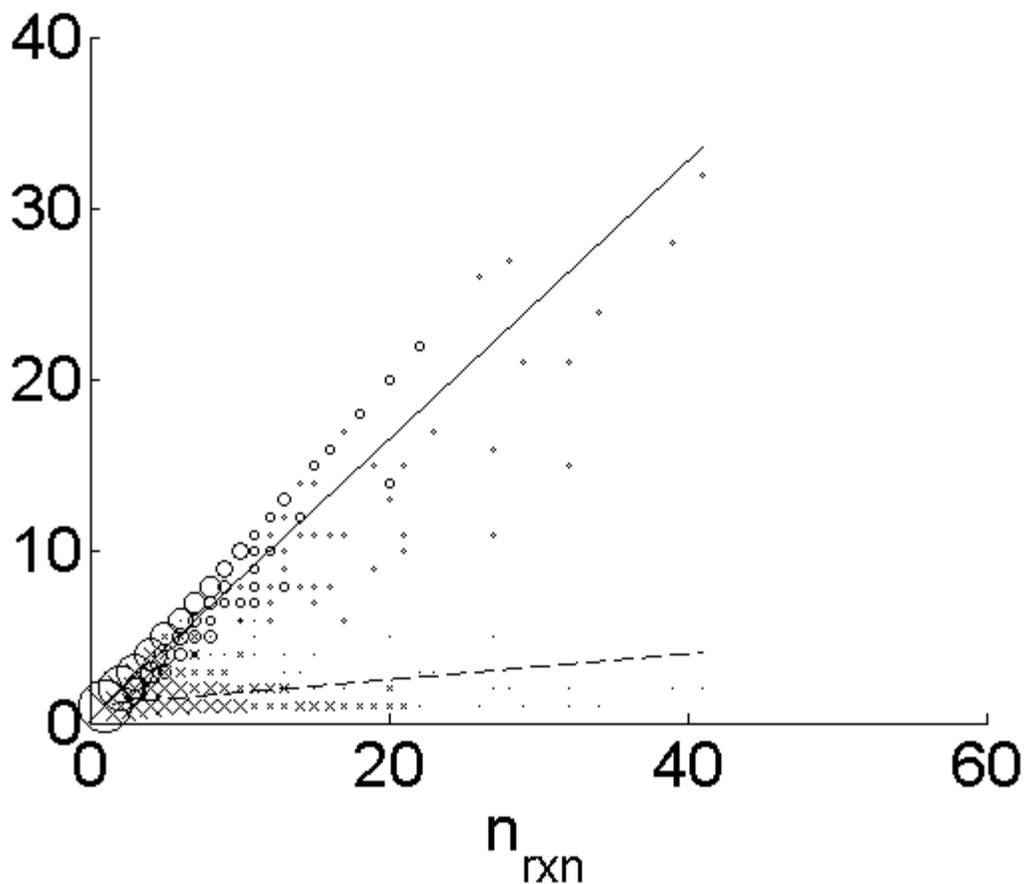

Figure 9. Comparison of the length of the pathways (circles and solid line) and the shortest distances of the targets from the core (crosses and dotted line).

As can be seen from Fig. 7, most of added pathways (around 97%) do not generate any byproducts. They only produce the intended target and $n_{\text{feedback}}$ metabolites in the core network of the organism to which they were added. The lack of byproducts indicates that pathways in our model satisfy the evolutionary constrains imposed on real-life organisms. Indeed, as previously proposed in Ref. it makes sense to assume that evolution favors pathways that achieve a given metabolic goal using the smallest number of enzymes and at the same time striving to generate the maximal possible yield. Unnecessary byproducts not only reduce the yield of the desired metabolic target, they also might become toxic in high concentrations and thus would require extra transporter proteins to pump them out.

## Discussion

Small world property of complex biomolecular networks has been extensively discussed in the literature during the last decade (see [9, 7, 8] for earliest reports in metabolic and protein interaction networks correspondingly). It was often assumed that the small world effect positively contributes to the robustness of the network by providing multiple redundant pathways for target production in metabolic networks or for propagation of signals along regulatory and protein interaction networks. In addition to its positive aspects the small world property in biomolecular networks also has a potentially negative side by facilitating system-wide propagation of undesirable cross-talk [10]. In the course of evolution different strategies appeared allowing organism to limit and attenuate these unwelcome side effects of global connectivity.

The extent of small world topology in metabolic networks has been recently disputed in [11]. There it was argued that many shortcuts in simple graph representations of metabolic networks are meaningless from biochemical standpoint. By taking into account additional structural information about metabolites Arita [11] dramatically increased the diameter of the metabolic network in *E. coli*. In our simulations of toolbox model we also encountered limitations of the simple graph representation giving rise to small world topology of metabolic networks. Small world by definition implies very short pathways (or equivalently overcritical network branching with exponentially growing lists of neighbors at distance $d$), which in its turn prevents the appearance of quadratic scaling in linear toolbox model.

How to reconcile this apparent contradiction? The answer known from pioneering studies of R. Heinrich and collaborators (see e.g. [6, 12, 13] ) is to altogether abandon the simple graph representation in favor of realistic treatment of multi-substrate reactions. A metabolic reaction with two or more substrates will not proceed at any rate until all these metabolites are present in the cell. This implicit "AND" function operating on inputs of multi-substrate metabolic reactions makes reaching a given metabolic target much harder task and ultimately leads to dramatically longer pathways (Fig. 9 quantifies this effect). These longer pathways in turn reinstate the quadratic scaling in the version of the toolbox model that was introduced in the previous chapter. One conclusion of our study is that the small world (overcritical) topology of the metabolic universe disappears in favor of the "large world", critical topology when multi-substrate reactions are properly taken into account. The increase in the effective diameter of the network due to this effect is dramatic. One goes from 3-4 steps diameter typical of a small world network of [8, 7] to ~8 steps of [11] and finally to 30-40 layers in the scope expansion process shown in Fig. 6 (see also Fig. 6 of [6]).

These arguments lead us to adapt the "scope expansion" algorithm by Heinrich et al [6] to pathway acquisition in the toolbox model. Not only it restored the "large world" properties such as quadratic scaling to the model, it also made the added pathways plausible from evolutionary standpoint. Unlike linear random walk pathways on KEGG network used in [2], pathways in the new version of the toolbox model have the smallest number of KEGG reactions to achieve their metabolic task (production of the target metabolite from the set of metabolites already present in organism's network). As can be seen in Fig. 7 a large fraction of these pathways also does not generate any byproducts. Accumulation of such byproducts inside a cell is potentially dangerous and would require specialized proteins to excrete them to the environment. The lack of byproducts also means that the useful yield of an added pathway is at or near its theoretical maximum. This all makes our model pathways likely endpoints of evolutionary optimization [5].

Optimality of metabolic pathways in central carbon metabolism was recently discussed in Ref. [14]. There it was shown that some (but not all) of these pathways coincide with the shortest walks in the space of possible metabolic transformations. This study also estimated a typical metabolic substrate can in principle be converted into any of the 20 different products in just one step. This quickly adds up to a very large number of biochemically feasible paths connecting metabolites to each other. However, this exponential growth does not necessarily result in a small world universal metabolic network. Indeed, evolutionary optimization leaves just a tiny fraction of these biochemically feasible reactions to be realized in any organism. The universal metabolic network formed by the union of all organism-specific metabolic networks is thus dramatically sparser than the set of all reactions allowed biochemistry. Thus, as demonstrated in Ref. [6] and the present study, the number of metabolites one could generate in N steps starting from a small core network and using KEGG-listed metabolic reactions instead of expanding as $20^N$ grows with N much more slowly (algebraically). The overall picture consistent with both our observations and those of Ref. [14] is that exponentially large, overcritical tree of all possible biochemical transformations is first pruned to an evolutionary optimized critical universal network out of which individual organisms select a subset of reactions necessary to accomplish their metabolic goals: that is to utilize nutrients in their environment and generate metabolic targets necessary for their operation.

Simplified "toy" models based on artificial chemistry reactions have a long history of being used to reveal fundamental organizational principles of metabolic networks:

- The recent model of Riehl et al [15] uses the simplest possible metabolites distinguished from each other only by the number of atoms of one element (e.g. carbon). All reactions in this case are of ligation/cleavage type (e.g. $2+3 \leftrightarrow 5$) constrained only by mass conservation. In spite of utmost simplicity of this artificial chemistry, the optimal pathways in this model display a surprisingly rich set of properties and bear some similarity to real-life metabolic pathways.

- The study of Pfeiffer et el [16] emphasizes the role of different chemical groups forming metabolites. They consider another artificial chemistry where metabolites are defined by binary strings indicating presence or absence of each of $N$ different chemical groups, and reactions transferring one such chemical group from one substrate which has it to another substrate which initially does not. Plausible evolutionary rules of their model give rise to complex scale-free metabolic networks emerging from the simple initial condition of $N$ completely non-specific transferase enzymes.

- Finally the artificial chemistry studied by Hintze et al [17] has molecules composed of three different types of atoms with different valences. Metabolites are linear molecules in which every atom is connected to others by as many bonds as specified by its valence. This model with rather complicated rules of evolution is then used to shed light on topics such as robustness and modularity of metabolic networks.

In our study we used the real-life (even if incomplete and sometimes noisy) metabolic universe of all reactions in the KEGG database. The only simplifying approximations remaining in the new most realistic version of toolbox model is random selection of metabolic targets in the course of evolution and easy availability of any subset of KEGG reactions for horizontal transfer. Both these approximations can be relaxed in later versions of the model. Another promising direction is to extend the toolbox model to artificial chemistry universal networks of Refs. [15], [16], [17]. While taking away from the realism of the model such extensions might help to broaden our intuition about what topological properties of the universal network determine the scaling properties of its species-specific subnetworks.

## Materials and Methods

The universal network used in our study consists of the union of all reactions listed in the KEGG database. The directionality of reactions and connected pairs of metabolites were inferred from the map version of the reaction formula: ftp.genome.jp/pub/kegg/ligand/reaction/reaction?mapformula.lst. The universal network with linearized pathways used to construct Figs. 2 and 3 consists of 1813 metabolites upstream of pyruvate and 2745 reaction edges out of which 1782 are irreversible and 963 are reversible. The metabolic network with branched and cyclic pathways used to construct Figs. 5-9 consists of 1861metabolites located downstream from the central metabolism and reachable from it by the scope expansion algorithm of Ref. [6]. It has 2819 reactions out of which 1402 are irreversible and the remaining 1417 are reversible. Table 1 shows the statistics for the number of substrates and products of these reactions. The list of core metabolites was obtained from KEGG Pathways Modules in the category "central carbohydrate metabolism" and extended with "currency" metabolites such as water, ATP and NAD. Simulations were done in Matlab and Octave.

Table 1A. Numbers of substrates and products of irreversible reactions.

| Irreversible reactions | | Number of products | | | | |
|---|---|---|---|---|---|---|
| | | 1 | 2 | 3 | 4 | 5 |
| Number of substrates | 1 | 157 | 141 | 4 | | |
| | 2 | 82 | 491 | 95 | 7 | |
| | 3 | 1 | 123 | 170 | 31 | 1 |
| | 4 | | 10 | 73 | 15 | |
| | 5 | | | 1 | | |

Table 1b. Number of metabolites at different ends of reversible reactions.

| Reversible reactions | | Numbers of substrates and products of reversible reactions | | | | |
|---|---|---|---|---|---|---|
| | | 1 | 2 | 3 | 4 | 5 |
| Number of metabolites in the other side of the reaction | 1 | 143 | 231 | 6 | | |
| | 2 | | 553 | 284 | 15 | |
| | 3 | | | 106 | 69 | 1 |
| | 4 | | | | 6 | 3 |

## Calculation of the average $\mu$ in the toolbox model on a critical tree

The total fraction of metabolites from the universal network that are present in an organism specific network is given by

$$\bar{\mu} = \sum_{d=1}^{d=d_{max}} N_M(d) \Big/ \sum_{d=1}^{d=d_{max}} N_M^{(U)}(d)$$
$$= \sum_{d=1}^{d=d_{max}} \mu(d) N_M^{(U)}(d) \Big/ \sum_{d=1}^{d=d_{max}} N_M^{(U)}(d).$$

The boundary condition at the last layer of the tree does not satisfy the Eq. (4) but instead is given by $\mu(d_{max}) = \tau$. One can easily show that for $d < d_{max}$ $\mu(d)$ rapidly (exponentially) converges to its steady state value $\mu = \sqrt{\tau}$ and stays at this level for as long as $d \gg 1$ when it starts rising again and ultimately approaches 1 at $d = 1$. In a large critical network the number of nodes in the last and the first several layers is small compared to the total number of nodes in the network. Hence in case of a critical network one has $\bar{\mu} \approx \mu$

## Solution to the toolbox model on a supercritical tree

On a random branching tree $\mu(d)$ - the fraction of nodes in the organism-specific network at distance $d$ from the root of the tree - satisfies the following difference equation:

$$\mu(d) = p\tau + k\mu(d+1) - (k-1+p)[\mu(d+1)]^2 \qquad \text{(S1)}$$

Here as before $p = p_0$ is the probability that a node is a leaf and $k = 1 - p_0 + p_2$ is the branching ratio that is 1 for a critical tree and >1 for an overcritical one. We are interested in small $\tau$ so that $\mu(d)$ and $\mu(d+1)$ are also small, and by keeping only the leading linear term in $\mu(d+1)$ one gets

$\mu(d) = p\tau + k\mu(d+1)$. The last layer $d_{max}$ is special since it contains only leaves and hence

$$\mu(d_{max}) = \tau$$

Iteratively solving Eq. (S1) one gets

$$\begin{aligned}\mu(d_{max} - l) &= p\tau + p\tau k + p\tau k^2 + \ldots + p\tau k^{l-1} + \tau k^l \quad \text{(S2)}\\ &= p\tau(k^l - 1)/(k-1) + \tau k^l \\ &= \tau k^l (k-1+p)/(k-1) - p\tau/(k-1) \\ &\approx \tau k^l / \mu_0 \end{aligned}$$

where $\mu_0 = (k-1)/(k-1+p)$. To arrive to this expression we have made an approximation by dropping the quadratic term in Eq. (S1). This made our estimation for $\mu(d)$ to increase without saturating at the steady state $\mu_0 = (k-1)/(k-1+p)$ derived from Eq. (6). To approximately take into account the effects of eventual saturation we assume that $\mu(d)$ follows the linearized difference equation until it reaches the steady state at the height $d_{max} - m$, and then $\mu(d)$ stays at a constant level $\mu_0$ over the region $d < d_{max} - m$. Solving the equation $\tau k^m / \mu_0 = \mu_0$ we get the depth of transition later after which $\mu(d)$ is at or near its steady state value:

$$m = \log_k (\mu_0^2 / \tau) \quad \text{(S3)}$$

Now we use the results of eq. (6) and (7) to calculate $n_M$, picking only the leading order:

$$\begin{aligned} N_M &= \sum_{l=0}^{m} N_M^{(U)}(d_{max}-l)\mu(d_{max}-l) + \sum_{l=m+1}^{d_{max}} N_M^{(U)}(d_{max}-l)\mu(d_{max}-l) \\ &= k^{d_{max}}\tau + \sum_{l=1}^{m}\left(k^{d_{max}}/k^l\right)\left(\tau k^l/\mu_0\right) + \sum_{l=m+1}^{d_{max}} \mu_0 k^{d_{max}}/k^l \\ &= k^{d_{max}}\tau + \left(\tau k^{d_{max}}/\mu_0\right)m + \mu_0 k^{d_{max}} \sum_{l=m+1}^{d_{max}} 1/k^l \\ &\approx \left(\tau k^{d_{max}}/\mu_0\right)\log_k\left(\mu_0^2/\tau\right) \approx N_M^{(U)}\frac{\tau}{\mu_0}\frac{k-1}{k}\log_k\left(\frac{\mu_0^2}{\tau}\right) \end{aligned} \qquad \text{(S4)}$$

Finally for $\mu = N_M/N_M^{(U)}$ one gets the Eq. (7)

$$\mu = \frac{\tau}{\mu_0}\frac{k-1}{k}\log_k\left(\frac{\mu_0}{\tau}\right) \qquad \text{(S5)}$$

The logarithmic correction $\log_k\left(\frac{\mu_0}{\tau}\right)$ comes from $m$ the depth of the transient layer. Unlike in a critical case considered before, in a supercritical tree the majority of all nodes are located in the last few layers and thus cannot be neglected. Moreover, when $\tau$ is small the depth of the transient layer $m$ is logarithmically large so we can safely drop other less important contributions to $N_M$ including deviations from the steady state in the other transient layer near $d=1$.